# BioFinBERT: Finetuning Large Language Models (LLMs) to Analyze Sentiment of Press Releases and Financial Text Around Inflection Points of Biotech Stocks

Valentina Aparicio[1,2,3], Daniel Gordon[1,2,3], Sebastian G. Huayamares[1,2,3], Yuhuai Luo[1,2,3]

[1]Quantitative & Computational Finance Program, Georgia Institute of Technology, GA, 30332, USA

[2]All four authors contributed equally.

[3]Correspondence: valentina_aparicio@hotmail.com, danielgordon643@gmail.com, sebastianhuayamaresmoreno@gmail.com, yuhuailuo0328@gmail.com

## Abstract

Large language models (LLMs) are deep learning algorithms being used to perform natural language processing tasks in various fields, from social sciences to finance and biomedical sciences. Developing and training a new LLM can be very computationally expensive, so it is becoming a common practice to take existing LLMs and finetune them with carefully curated datasets for desired applications in different fields. Here, we present BioFinBERT, a finetuned LLM to perform financial sentiment analysis of public text associated with stocks of companies in the biotechnology sector. The stocks of biotech companies developing highly innovative and risky therapeutic drugs tend to respond very positively or negatively upon a successful or failed clinical readout or regulatory approval of their drug, respectively. These clinical or regulatory results are disclosed by the biotech companies via press releases, which are followed by a significant stock response in many cases. In our attempt to design a LLM capable of analyzing the sentiment of these press releases, we first finetuned BioBERT, a biomedical language representation model designed for biomedical text mining, using financial textual databases. Our finetuned model, termed BioFinBERT, was then used to perform financial sentiment analysis of various biotech-related press releases and financial text around inflection points that significantly affected the price of biotech stocks.

## Keywords

Large language models, biotech public equity, financial sentiment analysis

# 1. Introduction

Small/mid-cap biotech companies are usually developing highly innovative and risky drugs and their stocks tend to respond very positively (sometimes 2X or higher returns) or negatively upon a successful or failed clinical readout or regulatory approval of their drug, respectively (**Figure 1**). These value-driving events are usually announced via press releases and disclosed in the 10Q quarter SEC filings of these biotech companies.

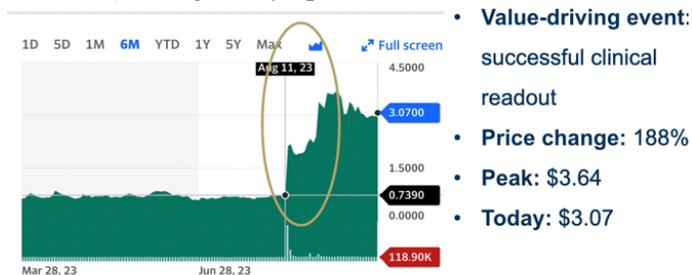

**Figure 1:** Sample value-driving event following a positive clinical readout by Taysha Gene Therapies ($TSHA).

Given the vast amount of recent biotech companies entering the public market in 2021 conducting clinical trials [1], we anticipate a large number of new therapies going through various clinical and regulatory value-driving events that can result in substantial moves in the biotech stock's value. Thus, we studied how various biotech stocks respond following the announcement of these clinical, regulatory, or other events through their press releases by applying large language models (LLMs) to these textual documents. Here, we applied and finetuned existing large language models LLMs to financial statements and press releases from biotech companies, perform sentiment analysis, and try to quantify and predict the corresponding biotech stock movements. We specifically performed our analysis on press releases of events announced between the years 2020 and 2023, given that there has been an increase in healthcare investments from the general public following the recent pandemic. Finally, we evaluated a trading strategy that yielded positive returns based on different holding periods for biotech stocks following positive events announced for their therapies in clinical development.

# 2. Overview

The goal of this work was to apply and finetune LLMs for sentiment analysis of financial statements and press releases to try to quantify and predict biotech stock movements. To do so, we broke our endeavors into three specific aims, as stated below.

**2.1. Aim 1:** Evaluate FinBERT on the press releases of an inflection point and 10Q statements before/after the inflection point, analyze the sentiment, and determine whether it predicted the stock move correctly.

|          | Large-cap  | Mid-cap    | Small-cap  | Micro-cap  |
|----------|------------|------------|------------|------------|
| Positive | > 1%       | > 2%       | > 3%       | > 4%       |
| Negative | < -1%      | < -2%      | < -3%      | < -4%      |
| Neutral  | -1% ~ 1%   | -2% ~ 2%   | -3% ~ 3%   | -4% ~ 4%   |

**Table 1:** Classification of positive, negative, and neutral value-driving events based on biotech stock price movements by market capitalization.

We gathered stock prices from Yahoo Finance and determined whether the event associated with the press release was (i) positive, corresponding to significant stock price upward change, (ii) neutral, corresponding to trivial or no stock price change, or (iii) negative, corresponding to stock price downward change. The definition of these thresholds varies by market capitalization for the stocks selected (**Table 1**).

**2.2. Aim 2:** Finetune and train BioBERT with existing financial text databases and use it to analyze the sentiment around the same press releases and financial statements.

Here, we assessed whether BioBERT finetuned and trained for financial sentiment analysis in addition to its pre-trained biomedical language model performed better than FinBERT, which has been trained exclusively on financial statements but has no training on biomedical language. Given that many of the press releases use clinical and biomedical terminology associated with a positive clinical readout that leads to a stock movement, we hypothesized that the finetuned BioBERT could perform better than FinBERT.

**2.3. Aim 3:** Propose and back-test a trading strategy around biotech stocks based on positive value-driving events.

To do this, we designed an equity strategy where different holding periods were evaluated for stocks bought and held following event-driving events. Given the differences in the behavior of stocks based on market capitalization, we evaluated the strategy accordingly.

## 3. Methods and Approach

### 3.1. Large Language Models (LLMs) Used

*FinBERT* [2, 3] is a pre-trained NLP model to analyze sentiment of financial text. It is built by further training the *BERT* large language model [4] in the finance domain, using a large financial corpus and thereby finetuning it for financial sentiment classification. *BioBERT* is a biomedical language representation model [5] designed for biomedical text mining tasks such as biomedical named entity recognition, relation extraction, and question answering. Both FinBERT and BioBERT rely on tokenization, a machine learning clustering algorithm employed in natural language processing.

### 3.2. Financial Sentiment Analysis

Sentiment analysis is the task of extracting sentiments or opinions of people or groups of people from written language [6]. Recent algorithmic and computational approaches for sentiment analysis can be segregated into two categories: (i) machine learning methods with features extracted from text via "word counting" [7, 8], and (ii) deep learning methods where text is represented by a sequence of embeddings [9, 10].

Importantly, while FinBERT was trained on pre-labeled financial textual datasets for sentiment analysis, BioBERT was only trained on biomedical corpora and terminology, but it was not designed for any form of sentiment analysis where the input is a sentiment score. Therefore, we first had to further train and finetune BioBERT to add financial sentiment analysis capabilities.

|  | Company | Ticker | MC ($M) |
|---|---|---|---|
| Large-cap: $10B - $200B | Moderna, Inc. | MRNA | $ 37,846.0 |
|  | BioNTech | BNTX | $ 24,725.0 |
|  | Alnylam Pharmaceuticals | ALNY | $ 22,108.0 |
|  | BioMarin Pharmaceutical | BMRN | $ 17,005.0 |
|  | Sarepta Therapeutics, Inc. | SRPT | $ 11,514.0 |
| Mid-cap: $2B - $10B | Ionis Pharmaceuticals | IONS | $ 6,755.0 |
|  | Ascendis Pharma | ASND | $ 5,289.0 |
|  | Apellis Pharmaceuticals | APLS | $ 4,572.0 |
|  | CRISPR Therapeutics | CRSP | $ 3,743.0 |
|  | Intellia Therapeutics | NTLA | $ 2,840.0 |
| Small-cap: $250M - $2B | Verve Therapeutics | VERV | $ 804.8 |
|  | Sana Biotechnology | SANA | $ 781.9 |
|  | Excientia | EXAI | $ 585.2 |
|  | Taysha Gene Therapies | TSHA | $ 555.3 |
| Micro-cap: < $250M | Atara Biotherapeutics | ATRA | $ 158.7 |

**Table 2:** Biotech stocks analyzed.

### 3.3. Finetuning of BioBERT for sentiment analysis

BioBERT was trained with existing financial textual databases and turned into a model capable of performing sentiment analysis of specific press releases where the biotech companies announce the value-driving event (clinical readout or regulatory approval, for instance) that generates a significant stock movement. To finetune BioBERT, we used existing pre-labeled financial textual databases, such as "financial_phrasebank" from HuggingFace.

### 3.4. Data Collection

We selected 26 biotech stocks of different market capitalizations (MC) with drugs in clinical development or recently approved that had various inflection points in the last few years (2020-2023). These will be classified into micro-cap: < $250M, small-cap: $250M-$2B, mid-cap: $2B-$10B, or large-cap: $10B-$200B (**Table 2**). First, we gathered the dates of various inflection events for each stock. Then, for each event, we obtained PDFs of the associated press releases with the clinical readout or regulatory approval that triggered the stock movement, as well as the SEC filings (10Qs and 10Ks for US based companies, and 6Ks and 20Fs for non-US based companies) of the quarters before and after these value-driving events. The PDFs of press releases, as well as the financial statements, were ran in FinBERT and then finetuned BioBERT.

For the trading strategy, stock prices were pulled from YahooFinance using the yfinance Python package. Evaluation of various holding periods were coded into Google Collaboratory, where we also back-tested the strategy on various subsets of the data. The Python code exported from the Google Collaboratory can be found here: https://bit.ly/BioFinBERT_arXivPreprint.

### 4. Results and Discussion

#### 4.1. BioFinBERT, a finetuned BioBERT for financial sentiment analysis

We first gathered PDFs of press releases from various biotech companies and the associated SEC filings released prior and posterior to the value-driving events. Diverse biotech companies were considered, encompassing different diseases being treated by different modalities (small molecules, biologics, gene therapies) and with different stages of clinical development (Phase 1, 2, or 3) or commercial stage drugs (**Table 2**). A total of 105 events were gathered and evaluated. These events were classified as (i) clinical events corresponding to clinical trial results readouts, (ii) regulatory events corresponding to FDA approvals or rejections of therapies investigated by the biotech companies, (iii) financial events corresponding to quarter earnings calls or announcements of follow-on financings, and (iii) other events such as commercial updates on sales or announcements of new collaborations with other biotech companies, acquisitions of smaller biotech companies, or participation in an upcoming conference (**Figure 2**).

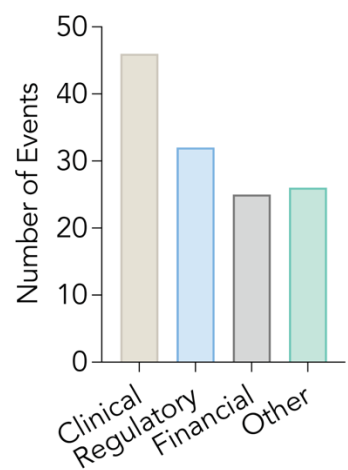

**Figure 2:** Types of events evaluated for biotech stocks.

When we pulled stock prices before and after these events and labeled the stock response to each press release associated with an event as detailed earlier for each market capitalization (**Table 1**), we found that, as expected, clinical events were most associated with positive stock movements (**Figure 3**). This is because companies tend to do big press releases mostly for positive clinical readouts. Negative readouts, however, are also disclosed either as press releases, which accounted for our negative clinical events, or in medical conferences. Regulatory events, on the other hand, had a more diverse combination of positive events from FDA approval notices, negative events from FDA rejection or safety concerns notices, or neutral events from FDA approvals or rejections that were already obviously anticipated based on recent clinical readout of a drug, so the corresponding price change was already built into the stock by the time the regulatory announcement happened. Financial events also had a fairly diverse combination of types of stock responses. Most of the positive stock movements resulting from financial events were associated with biotech companies exceeding projected sales of a commercial drug or having lower cash burn than projected, suggesting that their research and development had been optimized, lowering the company's odds of needing to raise follow-on funding rounds. Conversely, negative stock responses corresponded to financial events showing lower sales than projected or higher cash burns than anticipated. Finally, other events mostly corresponded to neutral or slightly positive stock movements. Some of the more positive stock responses from these other events included announcements of acquisitions or commercial collaborations or updates. For instance, a company expanding the sales of its vaccines in countries outside the US, or an acquisition of a smaller biotech company with promising therapeutics in early-stage development that would allow the acquiree to expand their drug pipeline.

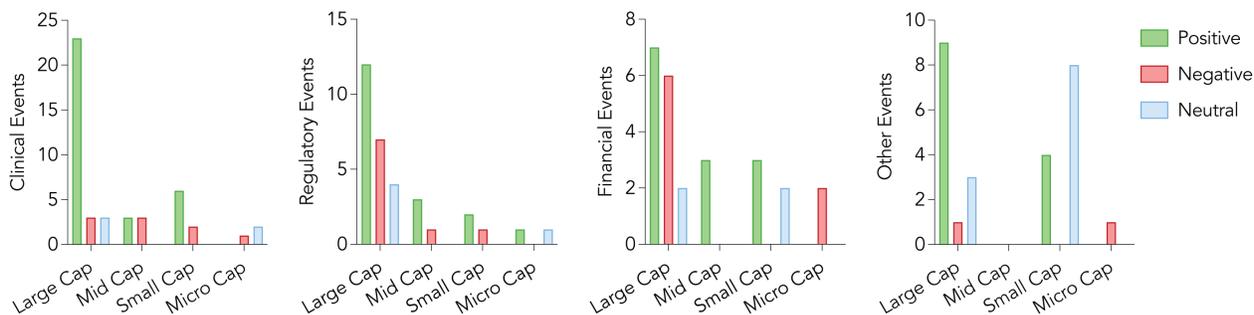

**Figure 3:** Classification of biotech stock responses (positive, negative, or neutral) by type of events (clinica, regulatory, financial, or other) and by market capitalization.

Given that FinBERT was already trained for sentiment analysis, we tested this model first on all press releases as well as prior and posterior financial reports. We defined a good prediction as a positive ($> 0$), negative ($< 0$), or neutral ($= 0$) sentiment score being attributed correctly to the press release associated with the event that resulted in a positive, negative, or neutral stock response, as defined earlier for different market capitalizations (**Table 1**). Similarly, if the prior or posterior financial report scores matched the label of the stock responses, we deemed it an accurate prediction. **Table 3** displays sample scoring of events from various biotech stocks. FinBERT predicted 66.86% (N=66/105) of the press releases accurately.

**Table 3:** Sample scoring for press releases and prior/posterior financial reports for $ATRA associated with positive, negative, or neutral value-driving events for biotech stocks.

| Ticker | Event Type | Date | Prior Price | Post. Price | Label | FinBERT | | | Fine-tuned BioBERT | | | Fine-Tuned BioBERT 2 | | |
|---|---|---|---|---|---|---|---|---|---|---|---|---|---|---|
| | | | | | | Event Score | Pre-Event 10Q Score | Event 10Q Score | Event Score | Pre-Event 10Q Score | Event 10Q Score | Event Score | Pre-Event 10Q Score | Event 10Q Score |
| ATRA | Regulatory | 2022-10-14 | $3.99 | $4.14 | Neutral | 0.263 | -0.106 | -0.036 | 0.237 | -0.159 | -0.069 | 0.158 | -0.029 | -0.007 |
| | Clinical | 2022-10-26 | $4.69 | $4.73 | Neutral | 0.216 | -0.106 | -0.036 | 0.405 | -0.159 | -0.069 | 0.081 | -0.029 | -0.007 |
| | Quarter Financial Results | 2022-11-09 | $4.90 | $3.88 | Negative | 0.385 | -0.106 | -0.036 | 0.192 | -0.159 | -0.069 | 0.115 | -0.029 | -0.007 |
| | Clinical | 2022-12-12 | $4.45 | $4.32 | Neutral | 0.186 | -0.106 | -0.036 | 0.163 | -0.159 | -0.069 | 0.070 | -0.029 | -0.007 |
| | Commercial | 2023-02-08 | $5.46 | $5.04 | Negative | 0.156 | -0.036 | -0.138 | 0.200 | -0.069 | -0.191 | 0.133 | -0.007 | -0.040 |
| | Quarter Financial Results | 2023-05-09 | $2.97 | $2.19 | Negative | 0.250 | -0.138 | -0.143 | 0.050 | -0.191 | -0.194 | 0.100 | -0.040 | -0.041 |
| | Regulatory | 2023-09-20 | $1.58 | $1.96 | Positive | 0.182 | -0.143 | -0.116 | 0.273 | -0.194 | -0.153 | 0.136 | -0.041 | -0.031 |
| | Clinical | 2023-11-09 | $1.21 | $0.24 | Negative | 0.148 | -0.116 | N/A | 0.111 | -0.153 | N/A | 0.148 | -0.031 | N/A |

Next, we finetuned BioBERT for financial sentiment analysis. To do this, we took three different approaches.

For our first attempt at finetuning BioBERT, we used the *financial_phrasebank* dataset from HuggingFace, which contains ~4,000 samples of financial news sentiment classification already attributed with positive, negative, or neutral labels. We used 3 epochs, which obtained the best results in the validation set of the data. Utilizing more than 10 epochs would result in significant overtraining of the model. We applied a learning rate of $2x10^{-5}$, based on trial and error. Ultimately, this finetuned BioBERT performed similarly to FinBERT for an F1 score of 0.9104, validation evaluation loss of 0.7088, and success rate of 62.86%. The model can be found in HuggingFace: https://huggingface.co/daniel-gordon/bioBERT-finetuned-financial-phrasebank

Our second attempt at finetuning BioBERT consisted of training it with a combined dataset of pre-labeled financial text for sentiment analysis. In addition to the *financial_phrasebank* from HuggingFace used earlier, we added the FiQA financial text sentiment dataset from Kaggle. Combined, these resulted in ~5,300 samples of financial news sentiment classification. For this second attempt, we used 5 epochs and a batch size of 16 for the training. A batch size of 32 yielded similar results on T4 GPU, so we retained a smaller batch size of 16. Our optimized learning rate for this attempt was $6x10^{-5}$; lower rates yielded slightly lower scores, and higher rates yielded significantly worse scores. Surprisingly, this attempt performed worse (validation F1 score = 0.8240, validation evaluation loss = 0.6959) than BioBERT trained only on the *financial_phrasebank* dataset in our first attempt. Thus, we did not move forward with this second approach.

Our third attempt was to curate our own dataset to combine with *financial_phrasebank* to finetune BioBERT. This dataset was obtained from 96 news articles from the biotech section of BioPharma-Dive news publisher. Our attempt was to create a similar dataset as *financial phrasebank* but using news articles specific to biotech companies. After segmenting and cleaning the news articles, we obtained ~2,000 sentences to label. We manually labeled 60% and used GPT-3.5 to pseudo-label the rest. The dataset can be found in HuggingFace: https://huggingface.co/datasets/daniel-gordon/biopharma-dive

We used the same training parameters to fine-tune the model, which can be found in HuggingFace: https://huggingface.co/daniel-gordon/bioBERT-finetuned-biopharma-dive

When comparing FinBERT with our finetuned BioBERT with *financial_phrasebank,* termed BioFinBERT, we saw overall similar prediction performances, although mild differences were detected when looking at different text data by type of event (**Figure 4**). First, we noticed that sentiment analysis using LLMs (FinBERT and BioFinBERT) on press releases associated with the value-driving events is more predictive than sentiment analysis on financial statements (10Qs, for example, either prior or posterior to the event) for biotech companies. This distinction was very significant across all types of events, and this is very particular to biotech stocks compared to stocks from other industries where single events do not drive the stock price up or down significantly.

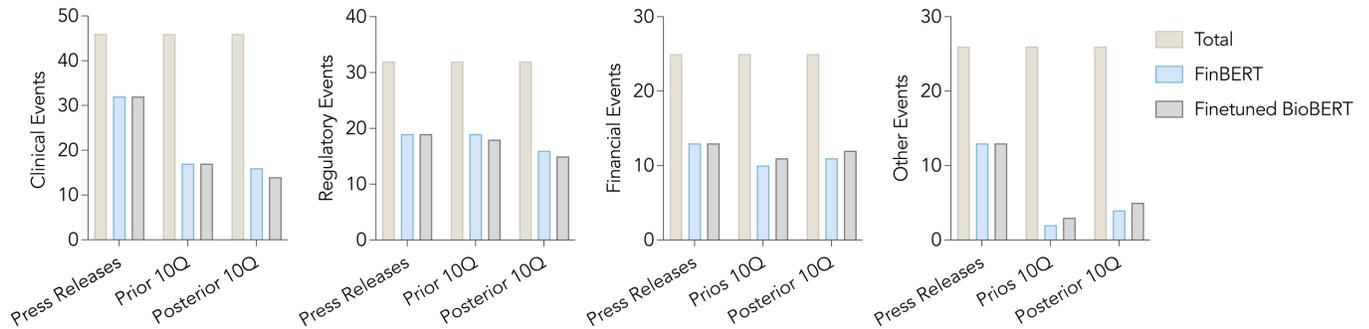

**Figure 4:** Classification of biotech stock responses (positive, negative, or neutral) by type of events (clinica, regulatory, financial, or other) and by market capitalization.

### 4.2. Biotech Sector Event-Driven Equity Trading Strategy

Finally, we developed an event-driven trading strategy around biotech-sector stocks. Our strategy executes market order based on the sentiment labels generated by BioFinBERT on various events including 10-Q, 10-K, 6-K, 20-F filings and press releases of biotech companies. If the sentimentGiven that biotech stocks display relatively larger volatility levels compared to other industries, we assessed different holding periods following positive events announced by the companies on the press releases studied with the LLMs. Specifically, our trading strategy implied holding biotech stocks following a positive event for 1, 2, or 3 months after the event's announcement. Our market benchmark was the NASDAQ Biotechnology Index (NBI) and the back-testing period was between 2020 and 2023. The evaluation metrics we assessed were cumulative returns, annualized returns, volatility, and annualized Sharpe ratio, and this was evaluated for stocks segregated into categories by market capitalization, consistent with all our analyses so far. **Table 4** reports the resulting metrics of our event-based equity strategy.

|  | Large | Mid | Small | Micro |
|---|---|---|---|---|
| Cumulative Return | -23% | 25% | 68% | 89% |
| Annualized Return | -6% | 6% | 14% | 17% |
| Volatility | 14% | 19% | 21% | 20% |
| Annualized Sharpe Ratio | -0.50 | 0.27 | 0.61 | 0.81 |

Holding Period: 1 Month

|  | Large | Mid | Small | Micro |
|---|---|---|---|---|
| Cumulative Return | 263% | -52% | 334% | 256% |
| Annualized Return | 38% | -17% | 44% | 37% |
| Volatility | 16% | 26% | 30% | 30% |
| Annualized Sharpe Ratio | 2.35 | -0.68 | 1.45 | 1.22 |

Holding Period: 2 Months

|  | Large | Mid | Small | Micro |
|---|---|---|---|---|
| Cumulative Return | 415% | -52% | 380% | 1414% |
| Annualized Return | 51% | -17% | 48% | 97% |
| Volatility | 29% | 26% | 29% | 224% |
| Annualized Sharpe Ratio | 1.71 | -0.68 | 1.60 | 0.43 |

Holding Period: 3 Months

**Table 4:** Summary of evaluation metrics of the event-based trading strategy proposed, classified by holding period and by market capitalization of biotech stocks.

For a 1-month holding period, holding micro-cap stocks following a positive event had the best performance, and holding large-cap stocks had the worst performance. This is possibly due to the fact that micro- and small-cap biotech stocks are value-driven meaning that the value of their clinical asset determines the value of their stock, while large-cap biotech stocks with commercial assets are no longer assessed based on the value of a single drug but also based on revenues generated from their commercial drugs, which can be responsible for the lower performance at a short holding period. When evaluating a 2-month holding period, large-cap biotech stocks had the best overall performance (highest cumulative return, lowest volatility, and highest Sharpe Ratio). This could be because given a longer period of time, these large-cap stocks have the chance of going through a quarter earnings report where they can have positive sales of commercial assets that are reflected in positive stock movements. Interestingly, small-cap biotech stocks were the best performing ones for the strategy with a 3-month holding period, with a 300% cumulative return and 1.31 annualized Sharpe Ratio, but a relatively low 31% volatility comparable to that of the large-cap biotech stocks. Overall, while potentially highly profitable, trading around biotech stocks is accompanied with large volatility.

## 4.3. Equity Event-Based Trading Strategy Based on Press Releases

The press releases documents provide a different kind of information than the 10-Qs and other financial statements. We found that the market reaction to Press Releases is more volatile because the topics and release dates aren't as consistent as financial statements. Therefore, we examined these over a wider range of 1 to 90 days of holding periods and three trading strategies:

1. The benchmark trading strategy consists in entering a position based on a 5% threshold difference between the Open and Close price on the release day, or the next trading day if the press release wasn't published on a trading day.
2. The "FinPhrase" strategy consists in using bioBERT fine-tuned only on *financial_phrasebank*.
3. The "BioPharma" strategy consists in using the bioBERT fine-tuned on *financial_phrasebank* dataset and on our curated *biopharma dive* dataset.

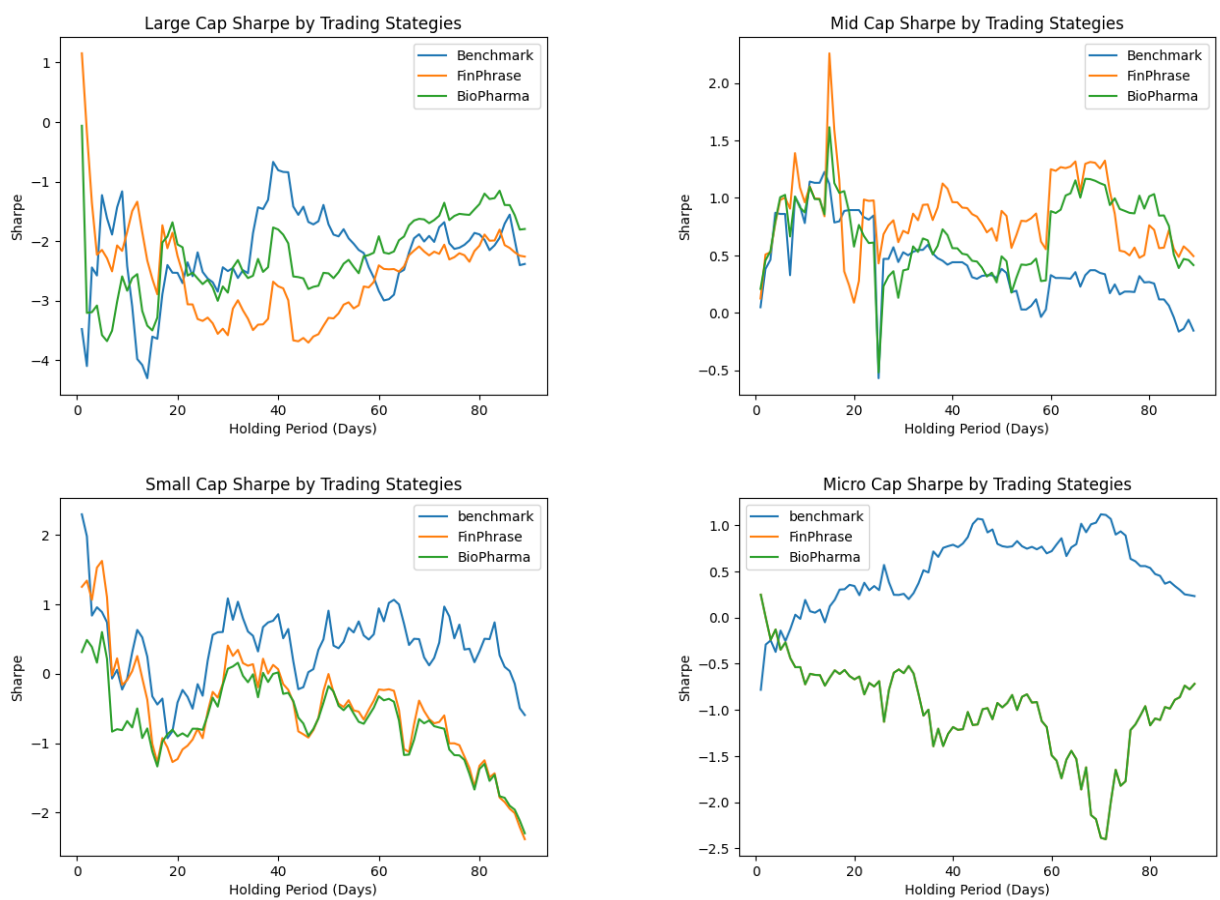

**Figure 5:** Annualized Sharpe Ratio of the "Benchmark", "FinPhrase", and "BioPharma" trading strategies for each market cap category.

When assessing the Annual Sharpe Ratio for each of the Trading Strategies in each market cap category (**Figure 5**), evaluated for 1 to 90 holding days in each case, we can draw several insights:

- BioPharma strategy works best for Large-Cap companies
- FinPhrase strategy works best for Mid-Cap companies

- Benchmark strategy is the best for Small- and Micro-Cap companies.

This suggests that the performance of trading strategies can vary a lot based on the Sentiment Classifier used to determine when to enter positions. And in some cases, a simple momentum strategy such as the benchmark can perform better. This leads us to believe that the curated dataset *biopharma dive* may not be as high quality as *financial_phrasebank* as we don't see significant improvement and in some cases the model leads to worse trading performance.

**5. Conclusions**

Biotech stocks can have significant price changes following single events of clinical, regulatory, financial, or commercial nature. We used LLMs to try to predict biotech stock price movements based on sentiment analysis on the press releases where these value-driving events were announced, as well as traditional financial statements corresponding to the quarters prior and posterior to these events. After finetuning BioBERT, an LLM trained on biomedical terminology, and adding sentiment analysis capabilities to it, we developed BioFinBERT, our finetuned LLM for performing financial sentiment analysis on biotech news. When tested against FInBERT, an LLM trained for general financial sentiment analysis, we saw that BioFinBERT performed similarly, with mild distinctions in press release perfomances. BioFinBERT can be tuned further to obtain increased performance and yield even better sentiment predictions of stock movements. Finally, we showed that an event-based trading strategy can be applied around these biotech stocks where we hold them following a positive event for different holding periods. Volatility and performance varied by market capitalization, as originally hypothesized.

**Author Contributions:** All four authors contributed equally.

**Funding.** This research did not receive any specific grant from funding agencies in the public, commercial, or not-for-profit sectors.

**Conflict of Interest:** The authors declare no conflict of interests.

7. Tripathy, A., A. Agrawal, and S.K. Rath, *Classification of sentiment reviews using n-gram machine learning approach.* Expert Systems with Applications, 2016. **57**: p. 117-126.
8. Whitelaw, C., N. Garg, and S. Argamon, *Using appraisal groups for sentiment analysis*, in *Proceedings of the 14th ACM international conference on Information and knowledge management*. 2005, Association for Computing Machinery: Bremen, Germany. p. 625–631.
9. Zhang, L., S. Wang, and B. Liu *Deep Learning for Sentiment Analysis : A Survey*. 2018. arXiv:1801.07883 DOI: 10.48550/arXiv.1801.07883.
10. Severyn, A. and A. Moschitti, *Twitter Sentiment Analysis with Deep Convolutional Neural Networks*, in *Proceedings of the 38th International ACM SIGIR Conference on Research and Development in Information Retrieval*. 2015, Association for Computing Machinery: Santiago, Chile. p. 959–962.
11